# Excellent magnetocaloric effect exhibited in the dense inorganic framework material Gd(OH)CO₃

**Yan-Cong Chen,**[a†] **Zhao-Sha Meng,**[a†] **Lei Qin**[b] **Yan-Zhen Zheng\*,**[b] **Jun-Liang Liu,**[a] **Fu-Sheng Guo,**[a] **Róbert Tarasenko,**[c] **Martin Orendáč,\*c Jan Prokleška,**[d] **Vladimír Sechovský,**[d] **and Ming-Liang Tong\*a**

**ABSTRACT:** The excellent magnetocaloric effect of a dense inorganic framework, orthorhombic Gd(OH)CO₃, is evaluated by isothermal magnetization and heat capacity measurements. The large $-\Delta S_m$, both in gravimetric and volumetric units, place it as a great candidate for cryogenic magnetic cooler.

## INTRODUCTION

The Magnetic refrigeration by utilizing the magnetocaloric effect (MCE) has been proposed to be an environmentally friendly and energy-efficient substitute of conventional gas compression techniques to acquire and maintain low temperature.[1] The MCE is referred to the isothermal magnetic entropy change ($\Delta S_m$) or the adiabatic temperature change ($\Delta T_{ad}$) when an external magnetic field is applied. Since the first observation in metallic Fe[2] and the following efforts to acquire ultra-low temperature by adiabatic demagnetization,[3-4] the MCE has been widely studied in a diversity of materials across the whole temperature range from micro-Kelvin to near room temperature.[5]

In the last decade, magnetic molecular clusters and coordination polymers have emerged exhibiting notable and promising MCE in the liquid helium temperature range,[6-9] showing the promising capability to replace the increasingly rare and expensive Helium-3.[10-17] The recent reports of Gadolinium(III)-based materials, due to its largest spin ($s = 7/2$), leads to significant enhancements in the MCE with the $\Delta S_m$ gradually catching up the commercial refrigerant Gd₃Ga₅O₁₂ (GGG) in the gravimetric unit of J kg⁻¹ K⁻¹.[18-21] However, there is still a long way to go since GGG holds great advantage from its high mass density of 7.08 g cm⁻³ to yield a huge $\Delta S_m$ in the volumetric unit of mJ cm⁻³ K⁻¹.

In fact, it has long been pointed out that the volumetric unit is more suitable to evaluate the performance of magnetic refrigerants to be placed in a designed refrigerator.[22-23] Therefore, the holy grail now comes to reduce the nonmagnetic components and increase the density while selecting suitable ligands to maintain the weak correlation between Gd³⁺ and low ordering temperature. In this regard, increasing the dimensionality of gadolinium-based materials has proven its effectiveness[19-21] and a recently reported dense Metal-Organic Framework, Gd(HCOO)₃, has exhibited a record MCE compared with other materials including GGG.[24]

As the ligand has now been simplified to formate and seems to reach an end, we have to ask a question whether the organic ligands are really essential in the pursuit of high performance magnetic refrigerants. Thus, here we report on the single-crystal structure and excellent magnetocaloric effect of an inorganic framework material, orthorhombic Gd(OH)CO₃. With the $-\Delta S_m$ up to 67.1 J kg⁻¹ K⁻¹ (359 mJ cm⁻³ K⁻¹) and $\Delta T_{ad}$ up to 24 K, our finding has placed it as a powerful competitor in cryogenic magnetic refrigeration.

## Experimental Section

Single crystal sample of Gd(OH)CO₃ was prepared by the hydrothermal reaction of GdCl₃·6H₂O (0.1 mmol), malononitrile (0.2 mmol) and deionized water (5 mL) under 180 °C for 72 hours followed by cooling to room temperature in air. Colorless crystals formed, and they were washed by deionized water and collected by filtration, yield c.a. 50% based on Gd³⁺. The phase purity was further confirmed by the powder XRD pattern **(Fig. 1)** IR (KBr cm⁻¹): 3459 (s), 2521 (w), 2340 (w), 1825 (s), 1790 (s). Anal. calcd. for CHGdO₄: C 5.13, H 0.43; found: C 5.31, H 0.56.

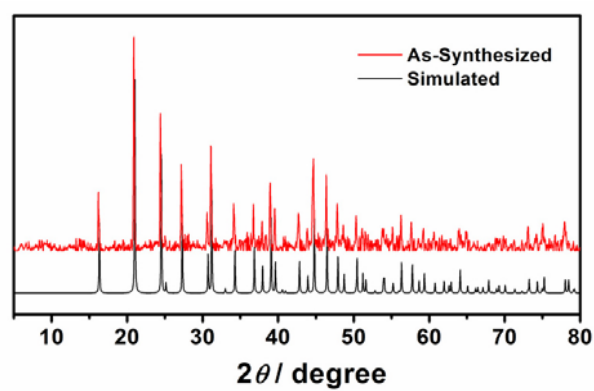

**Figure 1** Powder XRD pattern of Gd(OH)CO₃ compared to the simulation from single crystal structure.

Crystal diffraction data was recorded at 150(2) K on a Rigaku R-AXIS SPIDE Image Plate diffractometer with Mo Kα radiation, solved by direct methods and refined using SHELXTL program[25]. Crystal Data and Structural

Refinement are Listed in **Tables 1-2**. Further details on the crystal structure may be obtained from the Fachinformationszentrum Karlsruhe, 76344 Eggenstein-Leopoldshafen, Germany (fax: (+49) 7247-808-666; e-mail: crysdata@fiz-karlsruhe.de), by quoting the depository number CSD-426257 (Gd(OH)CO$_3$).

**Table 1** Crystal Data and Structural Refinement.

| Chemical formula | CHGdO$_4$ |
|---|---|
| Formula Mass | 234.27 |
| Crystal system | Orthorhombic |
| Space group | *Pnma* |
| Z | 4 |
| *a*/Å | 7.0770(7) |
| *b*/Å | 4.8730(9) |
| *c*/Å | 8.4353(6) |
| *α*/° | 90.00 |
| *β*/° | 90.00 |
| *γ*/° | 90.00 |
| Unit cell volume/Å$^3$ | 290.90(6) |
| Temperature/K | 150(2) |
| $\rho_{calcd}$ / g cm$^{-3}$ | 5.349 |
| $\mu$ (Mo K$\alpha$) / mm$^{-1}$ | 22.608 |
| No. of reflections measured | 2591 |
| No. of independent reflections | 356 |
| $R_{int}$ | 0.0305 |
| $R_1{}^a$ ($I > 2\sigma(I)$) | 0.0201 |
| wR$_2{}^b$ (all data) | 0.0538 |
| Goodness of fit on F$^2$ | 1.036 |

$^a R_1 = \Sigma||F_o| - |F_c||/\Sigma|F_o|$

$^b wR_2 = [\Sigma w(F_o{}^2 - F_c{}^2)^2/\Sigma w(F_o{}^2)^2]^{1/2}$.

**Table 2**. Selected Bond Lengths (Å) for Gd(OH)CO$_3$.$^a$

| Gd(1)-O(1) | 2.288(8) | Gd(1)-O(2)#4 | 2.515(5) |
|---|---|---|---|
| Gd(1)-O(1)#1 | 2.307(7) | Gd(1)-O(2)#5 | 2.555(5) |
| Gd(1)-O(3)#2 | 2.4955(17) | Gd(1)-O(2)#6 | 2.555(5) |
| Gd(1)-O(3) | 2.4955(17) | Gd(1)-O(2)#7 | 2.752(5) |
| Gd(1)-O(2)#3 | 2.515(5) | Gd(1)-O(2) | 2.752(5) |

$^a$Symmetry transformations used to generate equivalent atoms:

| #1 x-1/2,y,-z+1/2 | #2 x,y-1,z |
|---|---|
| #3 x+1/2,y,-z+1/2 | #4 x+1/2,-y-1/2,z+1/2 |
| #5 -x+1/2,-y,z+1/2 | #6 -x+1/2,y-1/2,z+1/2 |
| #7 x,-y-1/2,z | |

The magnetic measurements were performed on the polycrystalline samples using a Quantum Design MPMS XL-7 SQUID magnetometer. Low-temperature specific heat was studied on a Quantum Design PPMS with the $^3$He option adopting standard relaxation technique.

## Results and discussion

### Synthesis

Among the numerous gadolinium-based materials, the orthorhombic Gd(OH)CO$_3$ has be known for decades with its powder diffraction pattern listed in JCPDF as Gd$_2$O(CO$_3$)$_2$·H$_2$O, but the exact structure of it remains controversial since the single-crystal sample of this material is difficult to grow or correctly solved.[26-27] The frequently-used procedure employing one or the combination of carbonates, bicarbonates, hydroxides and urea always produce microcrystal or amorphous powder, and the unknown of the structure severely limits the deep research on this interesting material.

Our synthetic procedure of Gd(OH)CO$_3$ involved the hydrothermal reaction between Gd$^{3+}$ and malononitrile (NC-CH$_2$-CN). Though such a combination is not intended from the very beginning, the result shows gratifying, as large enough single crystals suitable for structural determination by X-Ray diffraction can be obtained. Indeed, it is not difficult to rationalized that the cyano groups can undergo hydrolysis into carboxylates plus amines in the hydrothermal condition, which is followed by decarboxylation and provide the crucial CO$_3{}^{2-}$ for the homogeneous precipitation of Gd(OH)CO$_3$. Meanwhile, the amines can act as buffering agent to prevent the rapid hydrolysis of Gd$^{3+}$ into Gd(OH)$_3$ or other indeterminable phase, thus also guarantee the good purity for subsequent property measurements.

### Crystal Structure

Single-crystal diffraction data analysis reveals the orthorhombic phase Gd(OH)CO$_3$ crystallizes in the *Pnma* space group and each asymmetric unit contains half of the chemical formula. In the crystal structure, each Gd$^{3+}$ ion is 10-coordinated to three chelating CO$_3{}^{2-}$, two mono-dentate CO$_3{}^{2-}$, and two OH$^-$ with the Gd…O bonds ranging from 2.29 to 2.75 Å (**Fig. 2a**). The CO$_3{}^{2-}$ groups are uniquely $\mu_5$-bridging with two $\mu_3$-O and one $\mu$-O between Gd$^{3+}$ (**Fig. 2b**).

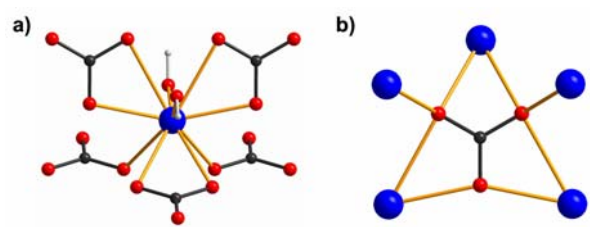

**Figure 2**. The coordination environment of a) Gd$^{3+}$ and b) CO$_3{}^{2-}$ Color Codes: Gd, blue; O, red; C, black; H, light gray.

As for the hydroxyl groups, they are $\mu$-bridging between Gd$^{3+}$, forming a series of 1-Dimensional zigzag chains along the *a* axis with the Gd…Gd distance of 3.82 Å

(**Fig. 3a**). However, the $CO_3^{2-}$ groups bridging with $Gd^{3+}$ turn out to form two groups of symmetry-related 2-Dimensional planes. In each group, the planes are parallel, while cross-group intersecting is established along the $b$ axis (**Fig. 3b**).

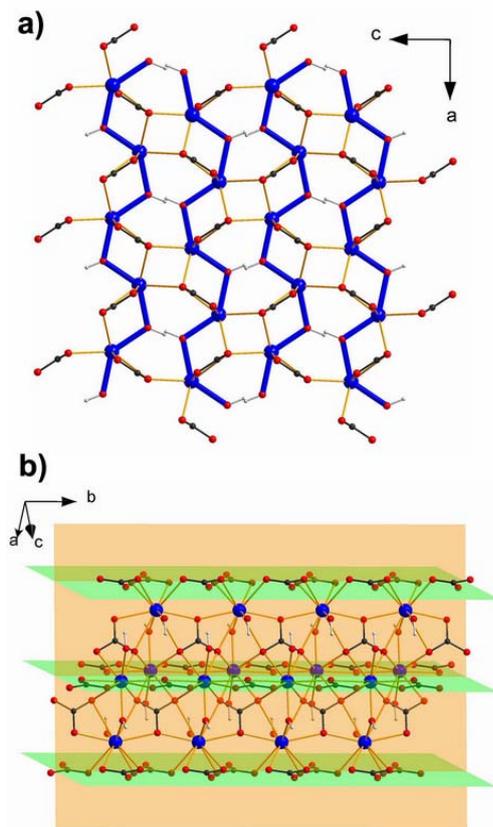

**Figure 3**. a) View along the $b$ axis of the crystal structure showing the 1-D zigzag $Gd^{3+}$-OH⁻ chain. b) View facing one of the 2-D $Gd^{3+}$-$CO_3^{2-}$ plane. Color Codes: Gd, blue; O, red; C, black; H, light gray

The structure of $Gd(OH)CO_3$ can now explain the high thermal stability of this material up to ~700 K[27] in contrast to the former assigning of $Gd_2O(CO_3)_2\cdot H_2O$. The dense inorganic framework structure without solvent-accessible porosity also gives rise to a large density of 5.349 g cm⁻³. Furthermore, we have found that the hydroxyl groups in the structure are μ-bridging rather than μ₃-bridging, which may not bring about strong magnetic coupling, and such a feature is extremely favorable for a cryogenic magnetic cooler.

## Magnetic Properties

Variable-temperature magnetic susceptibility measurement was performed on polycrystalline sample of $Gd(OH)CO_3$ in an applied dc field of 0.05 T (**Fig. 4**). At room temperature, the $\chi_m T$ value is 7.84 cm³ K mol⁻¹, in good agreement with the spin-only value expected for a free $Gd^{3+}$ ion with $g = 2$ (7.875 cm³ K mol⁻¹). Upon cooling, $\chi_m T$ stays essentially constant until approximately 30 K, followed by gradual decrease to the minimum value of 4.68 cm³ K mol⁻¹ at 1.8 K.

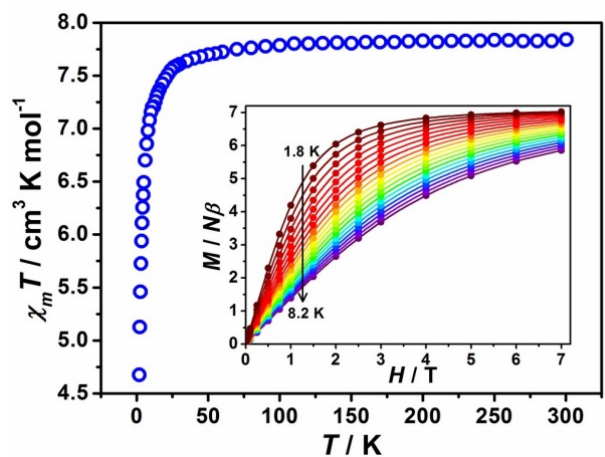

**Figure 4**. Temperature-dependencies of the magnetic susceptibility product $\chi_m T$ in 1.8 K ~ 300 K with the field of 0.05 T. inset: Magnetization versus field in the temperature range of 1.8 K ~ 8.2 K.

The inverse magnetic susceptibility ($1/\chi_m$, **Fig. 5**) obeys the Curie-Weiss law with $C$ = 7.86 cm³ K mol⁻¹, $\theta$ = -1.05 K from 300 K down to 1.8 K, indicating weak antiferromagnetic coupling. Surprisingly, despite the large proportion of OH⁻ in the polymeric structure, the overall magnetic coupling characterized by the Weiss constant $\theta$ is yet weaker than in many molecular clusters and cluster-organic frameworks comprising $Gd^{3+}$, which may arise from the difference in the bridging modes of OH⁻ (μ- rather than μ₃-).

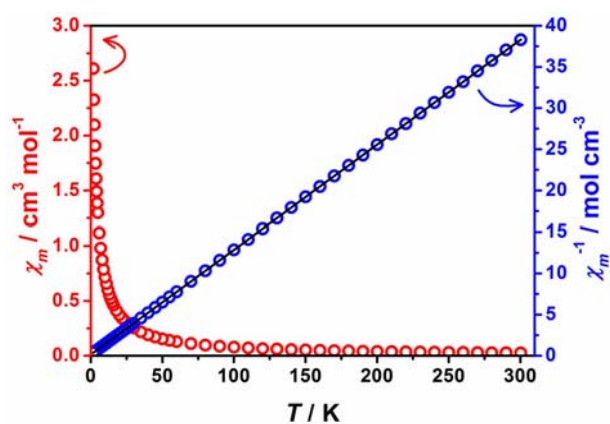

**Figure 5** Temperature-dependencies of the magnetic susceptibility ($\chi_m$) and inverse magnetic susceptibility ($1/\chi_m$). The temperature range is 1.8 K ~ 300 K and the applied field is 0.05 T. The black solid line represents the least-square fit for Curie-Weiss law.

The field-dependent magnetization at low temperatures were also measured in the temperature range 1.8 K ~ 8.2 K and field 0.025 T ~ 7 T (**inset of Fig. 4**). The magnetization increases steadily with the applied field and reach the saturation value of 7.0 $N\beta$ at 1.8 K, 7 T, in good agreement with the expected value of 7 $N\beta$ for a Gd$^{3+}$ ($s = \,^7/_2$, $g = 2$). The large magnetization values, together with the low molecular weight and high mass density, make this material promising candidate for cryogenic magnetic refrigeration, where the isothermal entropy change can be calculated by applying the Maxwell equation (**Fig. 6**):

$$\Delta S_m(T) = \int_0^H \left[\partial M(T, H) \big/ \partial T\right]_H dH$$

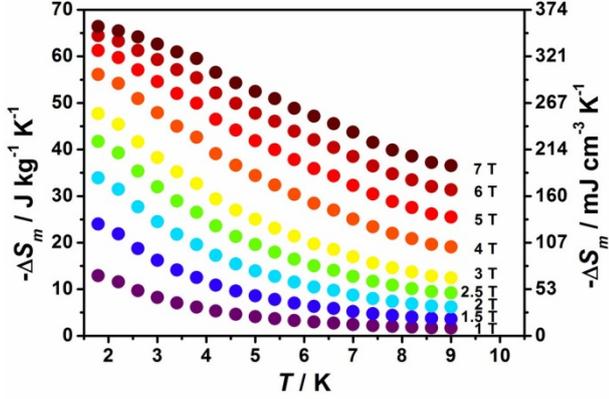

**Figure 6.** Temperature-dependencies of -$\Delta S_m$ obtained from magnetization. The data with applied field below 1 T are omitted for clarity.

In general, the -$\Delta S_m$ values grow gradually against reducing temperature, but rise progressively with increasing applied fields, reaching a maximum of 66.4 J kg$^{-1}$ K$^{-1}$ (355 mJ cm$^{-3}$ K$^{-1}$) at $T = 1.8$ K and $\Delta H = 7$ T, close to the theoretical limiting value of 73.8 J kg$^{-1}$ K$^{-1}$ (395 mJ cm$^{-3}$ K$^{-1}$) calculated from $R\ln(2s+1)/Mw$ with $s = \,^7/_2$ and $Mw = 234.3$ g mol$^{-1}$. The result here is already quite exciting; however the peak of -$\Delta S_m$ is still not reached in the aforementioned temperature, indicating the necessity of further investigation in the sub-Kelvin region.

**Heat Capacity Measurements**

To fully investigate the exciting MCE of this material, low temperature heat capacity ($C$) measurements were performed in the applied fields up to 9 T with the temperature down to ~0.5 K (**Fig. 7**). Clearly, the higher temperature region is dominated by lattice contribution arising from thermal vibration, which can be successfully fitted according to Debye's model and yield the Debye temperature ($\theta_D$) of 313(3) K with $r_D = 7$.[28-29] Such a high Debye temperature compared with other molecule-based materials is indicative of the stiffer frameworks consisting of strong chemical bonds and light ligand atoms. This is very important to yield a large $\Delta T_{ad}$ when the lattice vibration is forced to compensate for the variation of magnetic entropy in the adiabatic demagnetization process.

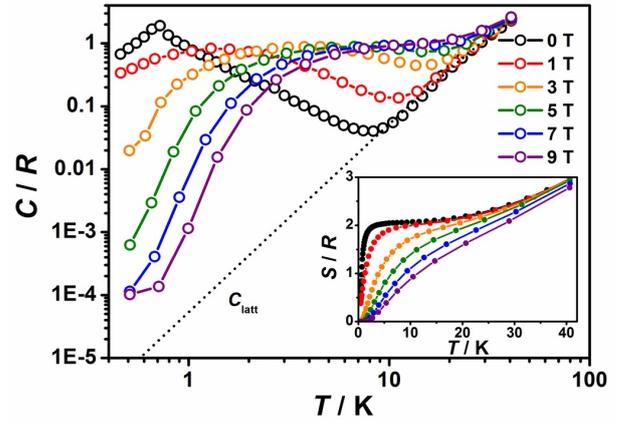

**Figure 7.** Temperature-dependencies of the heat capacity normalized to the gas constant in selected applied fields. The dotted line represents the lattice contribution. Inset: temperature-dependencies of the entropy obtained from heat capacity.

At lower temperatures, a significant domination of heat capacity by a field-dependent magnetic contribution can be found to show a broad Schottky type feature caused by the splitting of the $^8S_{7/2}$ multiplet. A small sharp anomaly is observed in the zero field at approximately 0.7 K but suppressed by applied fields, indicating the emergence of a phase transition. This can be attributed to the long-range antiferromagnetic interactions mediated by the polymeric network, which is further demonstrated by a downturn on the $\chi_m$-$T$ curve (**Fig. 8**).

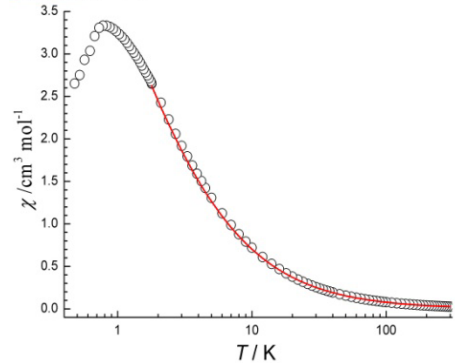

**Figure 8** Temperature-dependencies of the magnetic susceptibility ($\chi_m$) The temperature range is 0.46 K ~ 300 K and the applied field is 0.2 T. The red solid line represents the least-square fit for Curie-Weiss law.

Likely, such a behavior is also observed in the reported Gd(HCOO)$_3$ around 0.8 K,[24] however, the entropy content associated with the magnetic transition is quite low and it makes mere impact on the cooling capability as we can see below.

From the experimental $C$, the entropy can be obtained by numerical integration using:

$$S(T) = \int_0^T C(T)/T\, dT$$

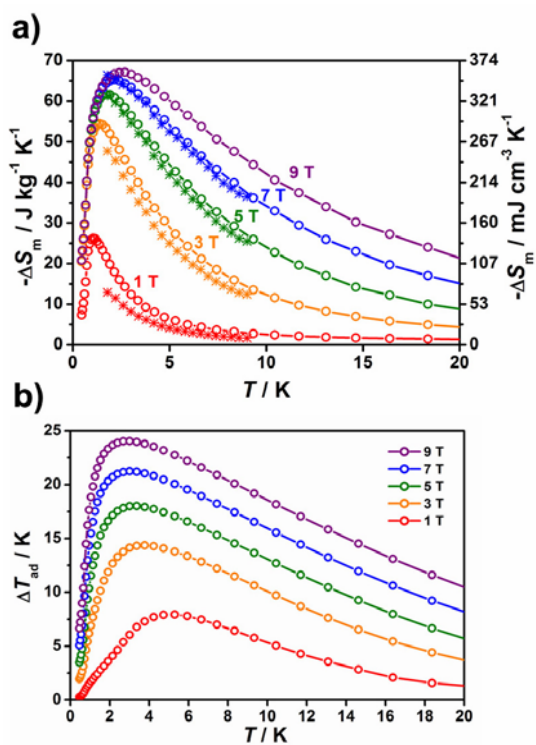

**Figure 9**. a) Temperature-dependencies of -$\Delta S_m$ obtained from magnetization (★) and heat capacity (●) for selected $\Delta H$. b) Temperature-dependencies of $\Delta T_{ad}$ for selected $\Delta H$.

Owing to the experimental inaccessibility of absolute zero, the experimental $C$ have to be extrapolated, and a constant value was added to the zero-field entropy for matching the high temperature saturation value of magnetic entropy ($S_{m, sat} = R\ln(2s+1) = 2.08 \ R$) as described in literature (**inset of Fig. 7**).[8,18] Then, the isothermal magnetic entropy change ($\Delta S_m$) and the adiabatic temperature change ($\Delta T_{ad}$) can be derived from the $S$-$T$ curves by vertical and horizontal subtraction, respectively (**Fig. 9**).[10]

With the information fully available, we can focus on the results of $\Delta S_m$ with easily recognizable peaks and the expected shift to higher temperature with increasing applied fields. Indeed, the performance for $\Delta H = 1$ T is already satisfactory with -$\Delta S_{m, max} = 26.2$ J kg$^{-1}$ K$^{-1}$ (140 mJ cm$^{-3}$ K$^{-1}$), and it increases sharply when larger fields are applied, namely 54.4 J kg$^{-1}$ K$^{-1}$ (291 mJ cm$^{-3}$ K$^{-1}$) for $\Delta H = 3$ T and finally reaches 67.1 J kg$^{-1}$ K$^{-1}$ (359 mJ cm$^{-3}$ K$^{-1}$) for $\Delta H = 9$ T.

Due to the advantages of extremely high magnetic density (Gd$^{3+}$ takes up 67% in the formula CHGdO$_4$.) and mass density (5.349 g cm$^{-3}$), the performance of Gd(OH)CO$_3$ not only surpass the reported molecular and molecule-based materials with the former record -$\Delta S_{m, max}$ of 50.1 kg$^{-1}$ K$^{-1}$ ([Mn$_3$Gd$_2$]$_n$)[17] and 125 mJ cm$^{-3}$ K$^{-1}$([Gd(HCO$_2$)(C$_8$H$_4$O$_4$)]$_n$),[21] but also exceed the recent groundbreaking Gd(HCOO)$_3$ (-$\Delta S_{m, max} = 215.7$ mJ cm$^{-3}$ K$^{-1}$).[24] Furthermore, since the employment of high magnetic field may not be suitable in all circumstances for the technical and economical consideration, the cooling capability under modest fields is also of great importance. Here for $\Delta H = 3$ T Gd(OH)CO$_3$ has already surpassed the commercial GGG whose -$\Delta S_{m, max} \approx 24$ J kg$^{-1}$ K$^{-1}$ (173 mJ cm$^{-3}$ K$^{-1}$) for the same applied field.[30] This prominent feature, along with the remarkable $\Delta S_m$ and $\Delta T_{ad}$ in higher applied field, further highlights the promising cooling power and energy efficiency of Gd(OH)CO$_3$. (**Fig. 10**).

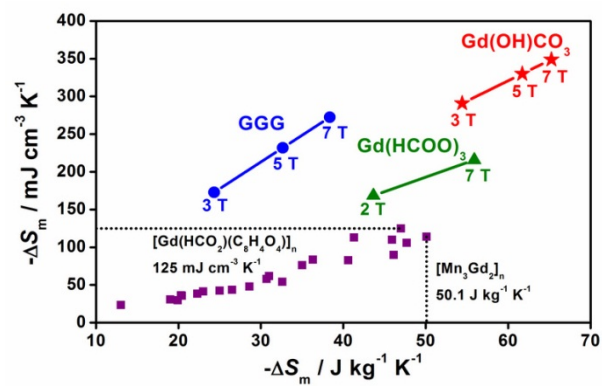

**Figure 10**. Comparison of the maximum -$\Delta S_m$ with selected $\Delta H$ for Gd(OH)CO$_3$(★), Gd(HCOO)$_3$ (▲), GGG (●) and the recently reported molecule-based magnetic refrigerants (■, where $\Delta H = 7$ T).

## CONCLUSIONS

In summary, we demonstrate the single-crystal structure and the experimental evaluation of the great magnetocaloric effect of an inorganic framework material, orthorhombic Gd(OH)CO$_3$. By magnetization and heat capacity measurements, it is found to be a great candidate for cryogenic magnetic refrigeration compared with the molecule-based materials and more importantly, the commercial GGG..

## ACKNOWLEDGMENT

This work was supported by the "973 Project" (2012CB821704), project NSFC (Grant no. 91122032, 90922009 and 21121061) and project APVV-0132-11. Part of the thermodynamic studies was performed in MLTL (http://mltl.eu/), which is supported within the program of Czech Research Infrastructures (project no. LM2011025).

## AUTHOR INFORMATION

[a] Key Laboratory of Bioinorganic and Synthetic Chemistry of Ministry of Education, State Key Laboratory of Optoelectronic Materials and Technologies, School of Chemistry & Chemical Engineering, Sun Yat-Sen University, Guangzhou, 510275, P. R. China.
E-mail: tongml@mail.sysu.edu.cn

[b] Center for Applied Chemical Research, Frontier Institute of Science and Technology, Xi'an Jiaotong University, Xi'an

710054，P. R. China.

E-mail: zheng.yanzhen@mail.xjtu.edu.cn

$^c$ Centre of Low Temperature Physics Faculty of Science, P.J. Šafárik University and Institute of Experimental Physics SAS, Park Angelinum 9, 041 54 Košice, Slovakia.

E-mail: martin.orendac@upjs.sk

$^d$ Faculty of Mathematics and Physics, Department of Condensed Matter Physics, Charles University, Ke Karlovu 5, CZ-12116 Prague 2, Czech Republic.

$^\dagger$These authors contributed equally to this work.